\documentclass[letterpaper,twocolumn,10pt]{article}
\usepackage{usenix}
\usepackage{epsfig}
\usepackage{subcaption} 

\usepackage{tabularx}
\usepackage{color}
\usepackage{booktabs}
\usepackage[utf8]{inputenc}
\usepackage[english]{babel}
\usepackage{graphicx}
\usepackage{amssymb}
\usepackage{url}
\usepackage{listings}
\usepackage{amsmath}
\numberwithin{equation}{section}
\usepackage{algorithm}
\usepackage{algpseudocode}
\usepackage{authblk}

\usepackage{xspace}
\usepackage{hyperref}

\usepackage[capitalise,noabbrev]{cleveref}

\usepackage{paralist}

\usepackage{titlesec}

\titlespacing{\section}{%
	0pt}{
	0.6\baselineskip}{
	5px}%
\titlespacing{\subsection}{%
	0pt}{
	0.6\baselineskip}{
	5px}%
\titlespacing{\subsubsection}{%
	0pt}{
	0.6\baselineskip}{
	5px}%
\titlespacing{\paragraph}{%
	0pt}{
	0.15\baselineskip}{
	8px}

\newcommand{\primeandprobe}{Prime+Probe\xspace}

\begin{document}
\title{Software Grand Exposure: SGX Cache Attacks Are Practical}





\author[1]{\rm Ferdinand Brasser}
\author[2]{\rm Urs M\"{u}ller}
\author[2]{\rm Alexandra Dmitrienko}
\author[2]{\rm Kari Kostiainen}
\author[2]{\rm Srdjan Capkun}
\author[1]{\rm Ahmad-Reza Sadeghi}
\affil[1]{System Security Lab, Technische Universit\"{a}t Darmstadt, Germany}
\affil[ ]{\{ferdinand.brasser,ahmad.sadeghi\}@trust.tu-darmstadt.de}
\affil[ ]{\vspace*{-1.5ex}}
\affil[2]{Institute of Information Security, ETH Zurich, Switzerland}
\affil[ ]{muurs@student.ethz.ch, \{alexandra.dmitrienko,kari.kostiainen,srdjan.capkun\}@inf.ethz.ch}

\date{}

\maketitle

\newcommand{\todo}[1]{\textcolor{red}{TODO: #1}}
\newcommand{\red}[1]{\textcolor{red}{#1}}

\begin{abstract}

Side-channel information leakage is a known limitation of SGX. Researchers have demonstrated that secret-dependent information can be extracted from enclave execution through page-fault access patterns. Consequently, various recent research efforts are actively seeking countermeasures to SGX side-channel attacks. It is widely assumed that SGX may be vulnerable to other side channels, such as cache access pattern monitoring, as well. However, prior to our work, the practicality and the extent of such information leakage was not studied. 

In this paper we demonstrate that cache-based attacks are indeed a serious threat to the confidentiality of SGX-protected programs. Our goal was to design an attack that is hard to mitigate using known defenses, and therefore we mount our attack without interrupting enclave execution. This approach has major technical challenges, since the existing cache monitoring techniques experience significant noise if the victim process is not interrupted. We designed and implemented novel attack techniques to reduce this noise by leveraging the capabilities of the privileged adversary. Our attacks are able to recover confidential information from SGX enclaves, which we illustrate in two example cases: extraction of an entire RSA-2048 key during RSA decryption, and detection of specific human genome sequences during genomic indexing. We show that our attacks are more effective than previous cache attacks and harder to mitigate than previous SGX side-channel attacks.  
 
\end{abstract}


\section{Introduction} \label{sec:intro}



Intel Software Guard Extension (SGX) \cite{Cos2016, ISCA2015} enables execution of security-critical application code, called enclaves, in isolation from the untrusted system software. Protections in the processor ensure that a malicious OS cannot directly read or modify enclave memory at runtime. Through a mechanism called sealing enclaves can encrypt and authenticate data for persistent storage. Processors are also equipped with certified keys that can issue remotely verifiable attestation statements on enclave software configuration. These SGX mechanisms (isolation, sealing, attestation) enable the development of applications and online services with improved security. The SGX architecture is especially useful in cloud computing applications. Data and computation can be outsourced to an external computing infrastructure without having to fully trust the cloud provider and the entire software stack.\footnote{Compared to other protection mechanisms SGX can provide significant advantages. Solutions based on special-purpose encryption offer limited functionality (e.g., searchable encryption \cite{bellare-crypto07}). Generic techniques (e.g., fully homomorphic encryption \cite{gentry-stoc09} and secure multi-party computation \cite{huang-usenix11}) are, for most applications, too slow. }

\paragraph{SGX information leakage.} However, previous research has demonstrated that SGX isolation has also weaknesses. The limited protected memory is used by unlimited number of enclaves, and therefore, memory management, including paging, is left to the OS \cite{Cos2016}. Consequently, the OS can force page faults at any point of enclave execution and from the requested pages learn secret-dependent enclave execution control flow \cite{Xu2015}. 

Information leakage is a serious concern, as it can defeat one of the main benefits of SGX -- the ability to compute over private data on an untrusted platform. Recent research has attempted to find ways to prevent such leakage. Currently the most promising system-level approach is to detect when the OS is intervening in enclave execution. For example, T-SGX~\cite{t-sgx} and D{\'e}j{\'a} Vu~\cite{incognito2017} detect page faults and allow the enclave to defend itself from a possible attack (i.e., to stop its execution). Sanctum \cite{sanctum} is an alternative security architecture, where the protected application itself is responsible for memory management, and thus able to prevent similar attacks.

Researchers \cite{Cos2016} and Intel \cite[p.~35]{intel-dev-guide-2016} have assumed that information may leak also through other side-channels such as caches that are shared between the enclave and the untrusted software. However, before our work, such leakage was not demonstrated and evaluated.

\paragraph{Our cache attack on SGX.} In this paper we demonstrate that SGX is indeed vulnerable to cache attacks. As a first use case we show our attack on the canonical RSA decryption and attack a standard sliding-window RSA implementation from the SGX SDK \cite{intel-exp-fast}. Using the \primeandprobe cache monitoring technique \cite{Osv2006, OsShTr2006} we can extract 70\% of the 2048-bit key with 300 repeated executions. From the extracted bits, the full RSA key can be effectively recovered \cite{boneh-98}. To the best of our knowledge, this work is the first to show that cache-based side-channel attacks are both practical and effective on SGX. 

Although cache-based attacks and cache monitoring techniques, such as \primeandprobe, are well studied, executing them in our setting involved a set of significant technical challenges. In particular, because our primary design goal was to explore attack techniques that cannot be easily mitigated by the recently suggested defensive approaches~\cite{t-sgx,incognito2017}, we opted to run both the victim and the attacker uninterrupted in parallel, so that the victim enclave is unaware of the attack and cannot take measures to defend itself. 
Hence, the victim cache monitoring needs to be fast, although monitoring all relevant cache sets can be slow. Furthermore, benign interrupts due to OS timers cause periodic enclave exists that cause severe interference in cache monitoring. Moreover, the execution of the victim itself can interfere with monitored cache sets. To overcome these challenges, we developed novel attack techniques. 
For instance, we leverage the capabilities of the privileged adversary to assign the victim process to a dedicated core, reduce the number of benign interrupts, and perform fast cache monitoring using CPU performance counters. Note that the SGX adversary model includes the capabilities of the OS. 

\paragraph{Current defenses.} Recent system-level defenses such as T-SGX~\cite{t-sgx} and D{\'e}j{\'a} Vu~\cite{incognito2017} can prevent those side-channel attacks that rely on frequent enclave interruption.
Our attack, however, does not require the interruption of the victim enclave and hence remains undetected by these defenses.
Besides system-level defenses, cache attacks can be tackled on the application level. Many cryptographic libraries provide encryption algorithm variants that have been specifically hardened against cache attacks. For every secret-dependent memory access the enclave can issue a set of memory accesses that manifest as changes in all the monitored cache sets. The accessed memory location is effectively hidden from the adversary. For instance, the NaCl library \cite{nacl} provides such side-channel resilient crypto implementations. Also the SGX SDK includes cryptographic algorithm variants that have been hardened against cache attacks \cite{intel-exp-secure}.

While such defenses can be effective, they require significant expertise and effort from the enclave developer. Assuming that every developer is aware of possible information leakage and able to harden his implementation against cache attacks is unrealistic. Automated tools that require no developer effort (e.g., oblivious execution \cite{maas-ccs13, liu-csf13, liu2015ghostrider} and ORAM \cite{stefanov-ccs13}) are difficult to deploy securely in SGX content and cause very high runtime overhead. Disabling caching is not practical either.

We argue that large classes of non-cryptographic SGX applications are vulnerable to cache attacks and illustrate this through our second use case, a genome indexing algorithm called PRIMEX~\cite{primex}, which uses hash tables to index a genome sequence. By monitoring the genome-dependent hash table accesses we can reliably identify if the processed human genome (DNA) includes a particular repeating sequence called microsatellite. Microsatellites are often used in applications such as forensics, genetic fingerprinting and kinship analysis \cite{Ballantyne-2010}.

We review known countermeasures and conclude that all of them have serious limitations, and none of them prevents our attacks effectively in practice.   


\paragraph{Contributions.} To summarize, this paper makes the following contributions:

\begin{compactitem}
	\item \textbf{Effective SGX cache attack.} We demonstrate that cache attacks are practical on SGX. Interestingly, our attack is more effective than previous comparable attacks. As part of our attack, we develop novel techniques to reduce side-channel noise.  

	\item \textbf{Leakage from non-cryptographic applications.} We show that non-cryptographic applications deployed within SGX are vulnerable to cache attacks. We demonstrate this through a case study on a genome analysis enclave.
	
	\item \textbf{Countermeasure analysis.} We show that none of the known defenses mitigates our attacks effectively in practice.
\end{compactitem}

The rest of this paper is organized as follows. In \cref{sec:background} we provide background information. \cref{sec:model} introduces the system and adversary model, and \cref{sec:design} explains our attack design. In \cref{sec:attack} we provide RSA decryption attack details and results. \cref{sec:genome} focuses on genomic enclave case study. We analyse countermeasures in \cref{sec:countermeasures}, discuss other algorithms and lessons learned in \cref{sec:discussion}, and review related work in \cref{sec:relwork}. \cref{sec:conclusion} concludes the paper.


\section{Background} \label{sec:background}
This section provides the necessary background for the rest of the paper.
We will start by describing Intel SGX, followed by a description of the cache architecture of current Intel x86 processors.
Afterwards we will introduce performance monitoring counters (PMC), a hardware feature that allows software to retrieve information about the state of hardware units.

\subsection{Intel SGX}
\label{sec:background:sgx}

SGX introduces a set of new CPU instructions for creating and managing isolated software components~\cite{Intel_SGX1,SGX_Ref}, called \emph{enclave}, that are isolated from all software running on the system including privileged software like the operating system (OS) and hypervisor.
SGX assumes the CPU itself to be the only trustworthy hardware component of the system, i.e., enclave data is handled in plain-text only \emph{inside} the CPU.
Data is stored unencrypted in the CPU's caches and registers, however, whenever data is moved out of the CPU, e.g., into the DRAM, it is encrypted and integrity protected.
This protects enclaves, for instance, from being attacked by malicious hardware components with direct memory access (DMA).

The OS, although untrusted, is responsible for creating and managing enclaves.
It allocates memory for the enclaves from a dedicated region of the physical memory called Enclave Page Cache (EPC).
It manages virtual to physical address translation for the enclave's memory and copies the initial data and code into the enclave. 
However, all actions of the OS are recorded securely by SGX and can be verified by an external party through (remote) attestation~\cite{Intel_SGX3}.
The sealing capability of SGX enables the persistent secure storage of enclave data, such that the data is only available to correctly created instances of one specific enclave.

During runtime of an enclave the OS can interrupt and resume the enclave like a normal process. 
Usually, upon an interrupt the OS is responsible for storing the current register content (context) of the interrupted process to free the register for use by the OS itself. 
To prevent information leakage, SGX handles the context saving of enclaves in hardware and erases the register content before passing control to the OS, called asynchronous enclave exit (AEX).
When an enclave is resumed, again the hardware is responsible for restoring the enclave's context, preventing manipulations.



\subsection{Cache Architecture}
\label{sec:cache}

In the following we will describe details of the Intel x86 cache architecture~\cite{Intel-manual,OptManual2012} required to understand the rest of the paper.
We focus on the cache architecture of the Intel Skylake processor generation, i.e., the type of CPU we used for our implementation and evaluation.\footnote{At the time of writing Intel SGX is only available on Intel Skylake and Kaby Lake CPUs, hence, only those two processor generations are relevant for this work. To the best of our knowledge there are no differences in the cache architecture between Skylake and Kaby Lake.}

Memory caching ``hides'' the latency of memory accesses to the system's dynamic random access memory (DRAM) by keeping a copy of currently processed data in cache.
When a memory operation is performed, the cache controller checks whether the requested data is already cached, and if so, the request is served from the cache, called a \emph{cache hit}, otherwise \emph{cache miss}. 
Due to higher cost (production, energy consumption), caches are orders of magnitude smaller than DRAM.
Hence, only a subset of the memory content can be present in the cache at any point in time.
The cache controller aims to maximize the cache hit rate by predicting which data are used next by the CPU core.
This prediction is based on the assumption of temporal and spatial locality of memory accesses.

\begin{figure}[t]
	\centering
	\includegraphics[width=0.75\linewidth]{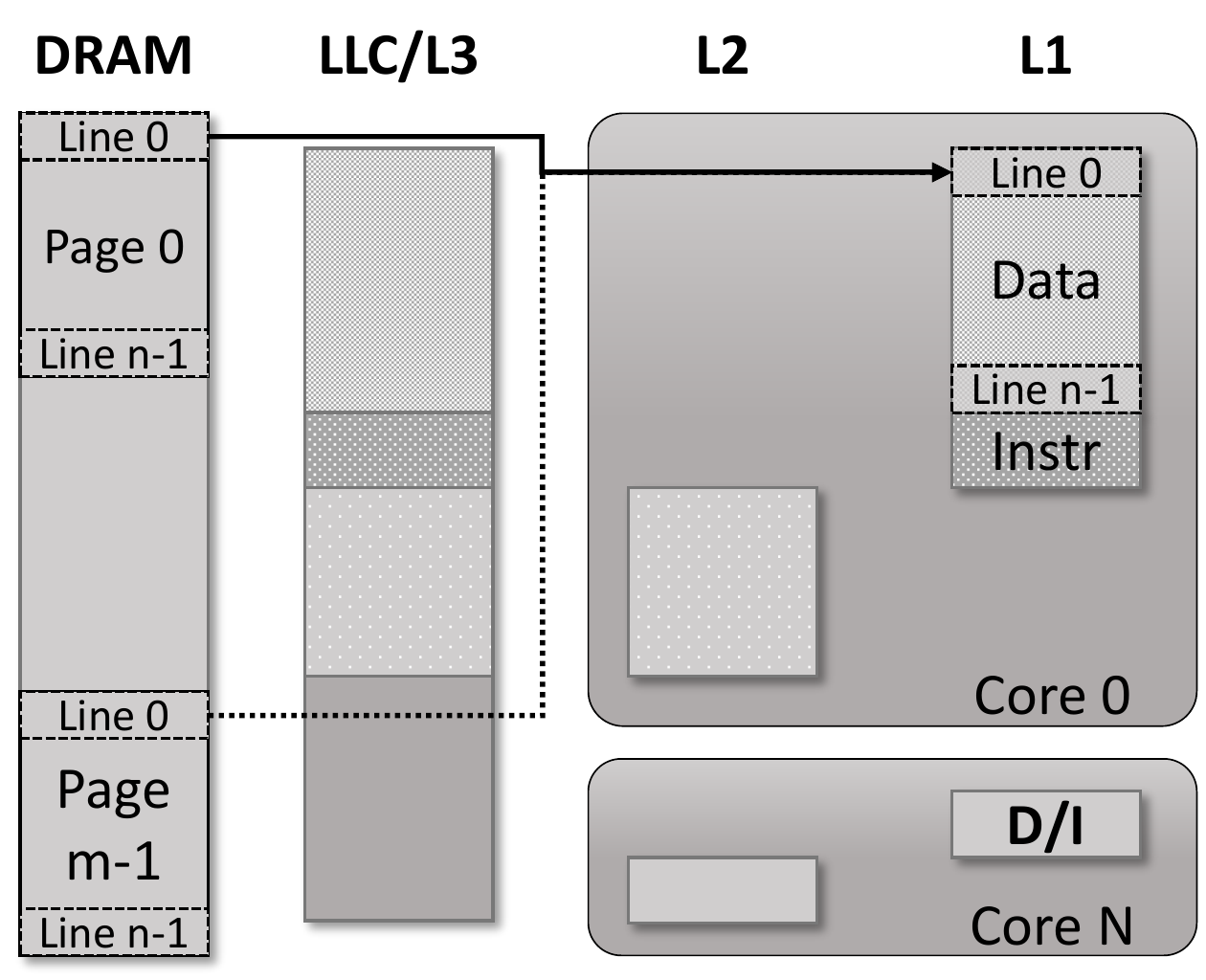}
	\caption{Cache hierarchy and configuration of Intel Skylake processors. The L3 cache is inclusive, i.e., all data stored in any per-core L1/L2 is also stored in L3. L1 cache is divided into separated parts for data and instructions.}
	\label{fig:cache_overview}
	\vspace*{-0.4cm}
\end{figure}

\cref{fig:cache_overview} shows the mapping of the main memory to the cache.
For each memory access the cache controller has to check if the data are present in the cache.
Sequentially iterating through the entire cache would be very expensive.
Therefore, the cache is divided into \emph{cache lines} and for each memory address the corresponding cache line can be quickly determined, the lower bits of a memory address select the cache line.
Hence, multiple memory addresses map to the same cache line, in \cref{fig:cache_overview} the first line of each \emph{cache page} in memory maps to the first cache line.
Having one cache entry per cache line quickly leads to conflicts, i.e., if memory from the first line of pages $0$ and $m-1$ are used at the same time, they conflict and the controller must evict data from a cache line to replace it with newly requested data.

The current Intel CPUs have a three level hierarchy of caches (\cref{fig:cache_overview}).
The last level cache (LLC), also known as level 3 (L3) cache, is the largest and slowest cache; it is shared between all CPU-cores.
Each CPU core has a dedicated L1 and L2 cache, but they are shared between the core's Simultaneous Multithreading (SMT) execution units (also known as hyper-threading).

A unique feature of the L1 cache is the separation into data and instruction cache. 
Code fetches only affect the instruction cache and leave the data cache unmodified, and the other way around for data accesses.
In L2 and L3 caches code memory and data memory compete for the available cache space.





\subsection{Performance Monitoring Counters}
\label{sec:background:pmc}

Performance Monitoring Counters (PMC) represent a feature of the CPU for recording hardware events.
Their primary goal is to give software developers insight into their program's effects on the hardware in order for them to optimize their programs.

The CPU has a set of PMCs, which can be configured to count different events, for instance, executed cycles, cache hits or cache misses for the different caches, mis-predicted branches, etc.
PMCs are configured by selecting the event to monitor as well as the mode of operation. 
This is done by writing to model specific registers (MSR), which can only be written with the \texttt{wrsmr} instruction (write to model specific register).
PMCs can only be set up by privileged software.
PMCs are read via the \texttt{RDPMC} instruction (read performance monitoring counters), which can be configured to be available in unprivileged mode.\footnote{``CR4.PCE -- Performance-monitoring counter enable. Enables execution of the RDPMC instruction at any protection level''~\cite{Intel-manual}.}

Hardware events recorded by PMCs could be misused as side-channels, e.g., to monitor cache hits or misses of a victim process or enclave.
Therefore, the SGX enclaves can disable PMCs on entry by activating a feature known as ``Anti Side-channel Interference'' (ASCI)~\cite{Intel-manual}.
This suppresses all thread-specific performance monitoring, except for fixed cycle counters.
Hence, hardware events triggered by an enclave cannot be monitored through the PMC feature.
For instance, cache misses of memory loaded by the enclave will not be recorded in the PMCs.



%
%


\section{System and Adversary Model}
\label{sec:model}
We assume a system equipped with Intel SGX, i.e., a hardware mechanism to isolate data and execution of a software component from the rest of the system's software that is considered untrusted.
The resources which are used to execute the isolated component (or enclave), however, are shared with the untrusted software on the system.

\begin{figure}[t]
	\centering
	\includegraphics[width=0.9\linewidth]{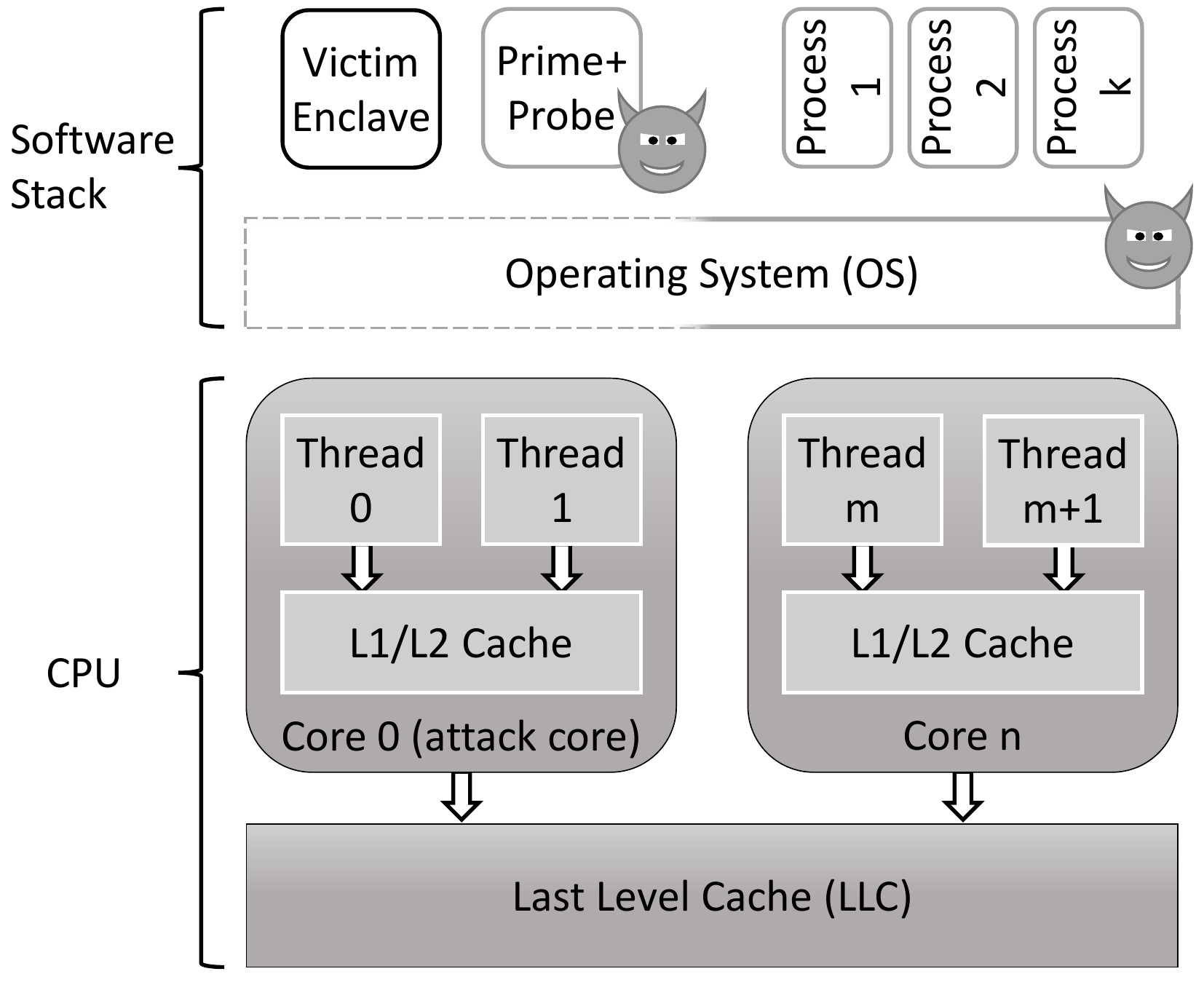}
	\caption{High-level view of our side channel attack; the victim enclave and the attacker's \primeandprobe code are running in parallel on a dedicated core. The attacker controlled OS ensures that no other code is executed on that core to minimize noise in its L1/L2 cache.}
	\label{fig:highlevel}
	\vspace*{-.4cm}
\end{figure}

The system's resources are managed by untrusted, privileged software.
In this work, we assume a system running Linux. For managing enclaves the system is relying on the Intel SGX software developer kit (SDK).
\cref{fig:highlevel} shows an abstract view of the adversary model, an enclave executing on a system with a compromised operating system, sharing a CPU core with an attacker process (\primeandprobe).

The adversary's objective is to learn secret information from the enclave, e.g., a secret key generated \emph{inside} the enclave through a hardware random number generator, or sensitive data supplied to the enclave \emph{after} initialization through a secure channel.
The attacker leverages his control over the system to minimize noise in the side channel.

\vspace{0.2cm}
\noindent\textbf{Adversary capabilities.}
The adversary is in control of all system software, except for the software executed inside the enclave.\footnote{Due to integrity verification, the adversary cannot modify the software executed inside the enclave, since SGX remote attestation would reveal tempering.}
Although the attacker cannot control the program inside the enclave, he does know the initial state of the enclave, i.e., the program code of the enclave and the initial data.
The attacker knows the mapping of memory addresses to cache lines and can reinitialize the enclave and replay inputs, hence, run the enclave arbitrarily often.
Further, since the adversary has control over the OS he controls the allocation of resources to the enclave, including the time of execution, and the processing unit (CPU core) the enclave is running on.
Similarly, the adversary can arbitrarily configure the system's hardware, e.g., define the system's behavior on interrupts, or set the frequency of timers.
However, the adversary cannot directly access the memory of an enclave. Moreover, he cannot retrieve the register state of an enclave, neither during the enclave's execution nor on interrupts.

\vspace{0.2cm}
\noindent\textbf{Attack scenarios.}
We consider two attack scenarios in this work (\cref{sec:attack} and \cref{sec:genome}).
The attacker knows the code and memory layout of the victim enclave, and hence knows memory locations accessed by the victim enclave.
The access pattern to the different memory locations allows him to draw conclusions about sensitive data processed by the victim.

For instance, a cryptographic algorithm uses precomputed data stored in different memory locations and accesses these values depending on the secret key.
The attacker observing the order of accesses to the precomputed values learns the key.
Similarly, an algorithm that inserts genomic data into a hash table allows the attacker to observe the insertion of genome sequences by monitoring which part of the tables are accessed. 
This allows the attack to detect subsequences within the genome that can be used, for instance, to identify persons.






\section{Our Attack Design} \label{sec:design}

Our attack is based on the \primeandprobe cache side-channel attack technique.
We will first explain the ``classical'' variant of \primeandprobe, then we discuss our improvements of that approach.

\subsection{\primeandprobe}
\label{sec:background:sidechannel}

All cache-based side-channel attacks are based on similar approaches.
The victim application and the attacker compete for the available cache, either by executing concurrently or interleaved.
The attacker aims to learn about the victim's cache usage by observing effects of the cache availability in its own program.
Different attack techniques have been developed that operate on different caches (L1 -- L3, instruction caches, virtual memory translation caches, etc.).

%
%
%
%

\begin{figure*}[th]
	\centering
	\includegraphics[width=.75\textwidth]{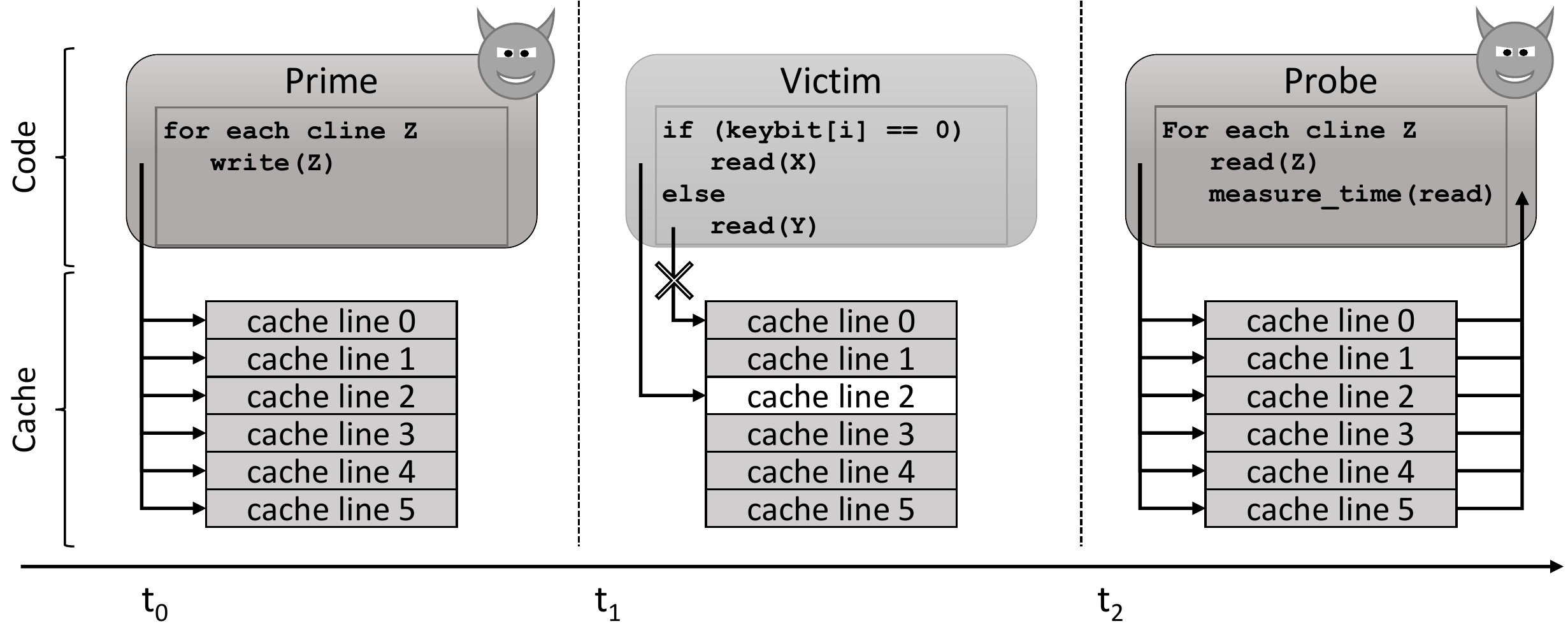}
	\vspace*{-.2cm}
	\caption{\primeandprobe side-channel attack technique; first the attacker primes the cache, next the victim executes and occupies some of the cache, afterwards the attacker probes to identify which cache lines have been used by the victim. This information allows the attacker to draw conclusion on secret data processed by the victim process.}
	\label{fig:prime_and_probe}
	\vspace*{-.2cm}
\end{figure*}

For our attack we adapted the \primeandprobe approach for learning about the victim's memory accesses, \cref{fig:prime_and_probe} shows the main steps.
First, the attacker \emph{primes} the cache, i.e., the attacker accesses memory such that the entire cache is filled with data of the attacker process.
At time $t_0$ the attacker writes to all cache lines, e.g., in current x86 CPU he writes to consecutive $4\,KB$ of memory.\footnote{To prime all cache sets the attacker needs to write to $\#cachesets$ cache pages, see \cref{sec:cache} for details.} 
Afterwards, at time $t_1$, the victim executes code with memory access that are dependent on the sensitive data processed by the victim.
In this example the victim processes a cryptographic key, which is sensitive data.
The victim accesses different memory locations depending on the currently processed key-bit.
In the example in \cref{fig:prime_and_probe} the key-bit is zero, therefore address $X$ is read.
Address $X$ is mapped to \emph{cache line 2}, hence, the data stored at $X$ are loaded into the cache and the data that were present before in that cache line gets evicted.
However, the data at address $Y$ are not accessed and therefore the data in \emph{cache line 0} remains unchanged.

At time $t_2$ the attacker probes which of his cache lines got evicted, i.e., which cache lines were used by the victim.
A common technique to check for cache line eviction is to measure access times: The attacker reads from memory mapped to each cache line and measures the access time. 
If the read operation returns the data fast, they were still cached, if the read operation takes longer, the data were evicted from the cache.
In the example in \cref{fig:prime_and_probe}, the attacker will observe an increased access time for \emph{cache line 2}.
Since the attacker knows the code and access pattern of the victim it knows that address $X$ of the victim maps to \emph{cache line 2}, and that the sensitive key-bit must be zero.
This cycle is repeated by the attacker for each sensitive key-bit that is processed by the victim and the attacker learns all bits of the key.

\subsection{\primeandprobe for SGX}

Extracting information through a side-channel is challenging due to noise.
The core idea of our attack is to reduce this noise.
We exploit the design of SGX where the OS (adversary) has control over the system configuration, and the scheduling and management of enclaves.

As mentioned before, we adapt the \primeandprobe approach to identify cache conflicts which we use as side-channel, i.e., we infer the victim's access to specific memory addresses based on the presence or absence of the corresponding entries in the cache.
To detect whether a cache line was used by the victim, the attacker accesses the same cache line and checks if his own cache entry was evicted, i.e., if the victim used that cache line.

To minimize the noise in the side-channel, we ensure that the cache is isolated and not affected by any system component except the victim enclave.
\cref{fig:highlevel} shows our approach to isolate the victim enclave on a dedicated CPU core, which only executes the victim and our attacker \primeandprobe code.
This way the per-core caches (L1/L2) are not influenced by any other process.
Furthermore, we need to ensure that the operating system itself does not pollute the cache of our \emph{attack core}.

\paragraph{Challenges.}
Reducing noise of the cache side-channel faces a number of technical challenges:

\begin{compactitem}
	\item[1.] Isolation of the attack core from use by other processes
	\item[2.] Minimization of cache pollution caused by the victim itself
	\item[3.] Running the victim uninterrupted to counter side-channel protection techniques and prevent cache pollution by the OS
	\item[4.] Reliably identify cache eviction caused by the victim
	\item[5.] Performing cache monitoring at a high frequency
\end{compactitem}

Below we will explain how we tackled each of these challenges.

\subsection{Noise Reduction Techniques}


\paragraph{(1.) Isolated attack core.}
By default Linux schedules all processes of a system to run on any available CPU core, hence, impacting all caches.
The attacker cannot distinguish between cache evictions caused by the victim and those caused by any other process.
Which process could cause the eviction is different based on whether considering the Last Level Cache (LLC) / Level~3 (L3)\footnote{The LLC is synonymic to the Level~3 (L3) cache in current processors.} or the Level~1 or Level~2 (L1/L2) cache.
By modifying the Linux scheduler, the adversary can make sure that one core (we call it \emph{attacker core}) is exclusively used by the victim and the attacker (``Core 0'' in \cref{fig:highlevel}).
This way no other process can pollute this core's L1/L2 cache.

Beside other processes, the OS can pollute the cache as well, we discuss this challenge below.

\paragraph{(2.) Self-pollution.}
The attacker needs to observe specific cache lines that correspond to memory locations relevant for the attack.
From the attacker's point of view it is undesirable if those cache lines are used by the victim for any other reason than accessing these specific memory locations, e.g., by accessing unrelated data or code that map to the same cache line.

In our attack we use the L1 cache.
It has the advantage of being divided into a data cache (L1D) and an instruction cache (L1I).
Therefore, code accesses, regardless of the memory location of the code, never map to the cache lines of interest to the attacker.
Victim accesses to unrelated data mapping to relevant cache lines leads to noise in the side-channel.

\paragraph{(3.) Uninterrupted execution.}
Interrupting the victim enclave yields two relevant problems. (1) When an enclave is interrupted, an asynchronous enclave exit (AEX) is performed and the operating system's interrupt service routine (ISR) in invoked (see \cref{sec:background:sgx}).
Both, the AEX and the ISR use the cache, and hence, induce noise into it. 
(2) By means of transactional memory accesses an enclave can detect that it has been interrupted.
This feature has been used for a side-channel defense mechanism~\cite{t-sgx,incognito2017}.
We discuss the details in \cref{sec:countermeasures}.
Hence, making the enclave execute uninterrupted ensures that the enclave remains unaware of the side-channel attack.

In order to monitor the changes in the victim's cache throughout the execution, we need to access the cache of the attack core in parallel.
For this we execute the attacker code on the same core. The victim is running on the first SMT (Simultaneous Multithreading) execution unit while the attacker is running on the second SMT execution unit (see \cref{fig:highlevel}).
As the victim and attacker code compete for the L1 cache, the attacker can observe the victims effect on the cache.

The attacker code is, like the victim code, executed uninterrupted by the OS.
Interrupts usually occur at a high frequency, e.g., due to arriving network packages, user input, etc.
By default interrupts are handled by all available CPU cores, including the attack core, and thus the victim and attacker code are likely to be interrupted.
The OS code executed on arrival of an interrupt will pollute the cache, or the victim enclave could detect its interruption, assume an attack, and stop itself.

To overcome this problem we configured the interrupt controller such that interrupts are not delivered to the attack core, i.e., it can run uninterrupted.
The only exception is the timer interrupt which is delivered per-core.
Each CPU core has a dedicated timer and the interrupt generated by the timer can only be handled by the associated core.
However, we reduced the interrupt frequency of the timer to $100\,Hz$, which allows victim and attacker code to run for $10\,ms$ uninterrupted.
This time frame is sufficiently large to run the complete attack undisturbed (with high probability).\footnote{When an interrupt occurs, by chance, during the attack phase the run can be repeated. If the attack phase is longer than $10\,ms$ the timer frequency can be reduced further.}
As a result, the OS is not executed on the attack core during the attack in progress, which is shown by the dashed-line OS-box above the attack core in \cref{fig:highlevel}.
Also, the victim is not interrupted, thus, it remains unaware of the attack.



\paragraph{(4.) Monitoring cache evictions.}
In the previous \primeandprobe attacks, the attacker determines the eviction of a cache line by measuring the time required for accessing memory that maps to that cache line.
This timing based measurements represent an additional source of noise to the side-channel.
Distinguishing between cache hit and miss requires precise time measurements, for instance for the L1 cache a cache hit takes at least $4~cycles$.
If the data got evicted from L1 cache, they can still be present in the L2 cache.
In this case, when the data are accessed, they will be read from L2 cache, which takes $12~cycles$ in the best case.\footnote{Reported values for Skylake architecture, however, ``Software-visible latency will vary depending on access patterns and other factors''~\cite{OptManual2012}.}
This small difference in access times makes it challenging to distinguish a cache hit in L1 cache and a cache miss in L1 that is served from L2 cache.
Reading the time stamp counter, to determine the access time, by itself suffers from noise in the order the effect to be observed.
Thus, when the timing measurement does not allow for a definitive distinction between a cache hit and a cache miss, the observation has to be discarded.
To eliminate this noise we use existing Performance Monitoring Counters (PMC) to determine if a cache line got evicted by the victim.
This is possible in the SGX adversary model because the attacker controls the OS and can freely configure and use the PMCs.

The intuitive approach to monitor cache related events of the victim are prevented by the fact that PMCs are disabled for enclave code (cf. \cref{sec:background:pmc}).
However, the attacker's \primeandprobe code shares the cache with the victim.
The attacker has primed all cache before the victim is executed.
Next the victim executes and evicts a subset of cache lines.
Hence, when the attacker probes the cache these lines will result in a cache miss.
The attacker uses PMC to identify these cache misses, learning which cache lines were used by the victim.


\paragraph{(5.) Monitoring frequency.}
As discussed before, the victim should run uninterrupted while its cache accesses are monitored in parallel.
Hence, we need to execute priming and probing of the cache at a high frequency to not miss relevant cache events.
In particular, probing each cache line to decide whether it has been evicted by the victim is time consuming and leads to a reduced sampling rate.
The required monitoring frequency depends on the frequency at which the victim is accessing the secret-dependent memory locations.
To not miss any access the attacker has to complete one prime and probe cycle before the next access occurs.
In our implementation the access to PMCs is the most expensive operation in the \primeandprobe cycle.

To tackle this challenge we monitor individual (or a small subset of) the cache lines over the course of multiple executions of the victim.
In the first run we learn the victim's accesses to the first cache line, in the second run accesses to the second cache line, and so on.
By aligning the results of all runs we learn the complete cache access pattern of the victim.

%

\section{RSA Decryption Attack} 
\label{sec:attack}

\renewcommand{\algorithmicrequire}{\textbf{Input: }}
\renewcommand{\algorithmicensure}{\textbf{Output: }}

In this section we describe how we apply the above attack techniques to the canonical example of key recovery from RSA decryption. We first describe our victim algorithm and implementation, then our attack details, and finally the key extraction results.

\subsection{Victim Enclave}
\label{subsec:victim}

As our victim enclave we chose an RSA implementation from the Intel IIP crypto library in the Intel SGX SDK. The attacked decryption variant is a fixed-size sliding window exponentiation, the code is available online at \cite{intel-exp-fast}. The Intel IIP library includes also a variant of RSA that is hardened against cache attacks~\cite{intel-exp-secure}. We discuss such defenses and their limitations in \cref{sec:countermeasures}. In this section we focus on demonstrating how effective our attack techniques can be against standard cryptographic implementations. 

The chosen decryption algorithm uses the Chinese Remainder Theorem (CRT) optimization, where two values $d_p$ and $d_q$ are pre-computed from the private key primes $p$ and $q$. To decrypt a message, separate exponentiation operations are performed using $d_p$ and $d_q$. For our experiments we use an RSA key size of 2048 bits which means that the decryption performs two 1024-bit exponentiations. 

\begin{algorithm}[th]
    \caption{Fixed-window exponentiation}
    \label{alg:exp}

    \algorithmicrequire{$a, e, N \in \mathbb{N}$}
    \newline
    \algorithmicensure{$x \gets {a}^{e} \mod N$}

    \begin{algorithmic}[1]

    \State Precompute $g[i] \gets a^i$ for $1 \leq i \leq 2^k$
	\State Let $e = (e_j, e_{j-1}, \ldots, e_1, e_0)$ be the base $2^k$ representation
		   of the exponent $e$ with $e_j \neq 0$

	\State Initialize $x \gets e_j$
	\For {$i \gets j - 1$ \textbf{ down to } 0}
		\State $x \gets x^{2^k} \mod N$
		\If {$e_i \neq 0$}
			\State $x \gets g[e_i] \cdot x \mod N$ 
		\EndIf
	\EndFor
    \end{algorithmic}
\end{algorithm}

\begin{figure}[t]
	\centering
	\includegraphics[width=0.9\linewidth]{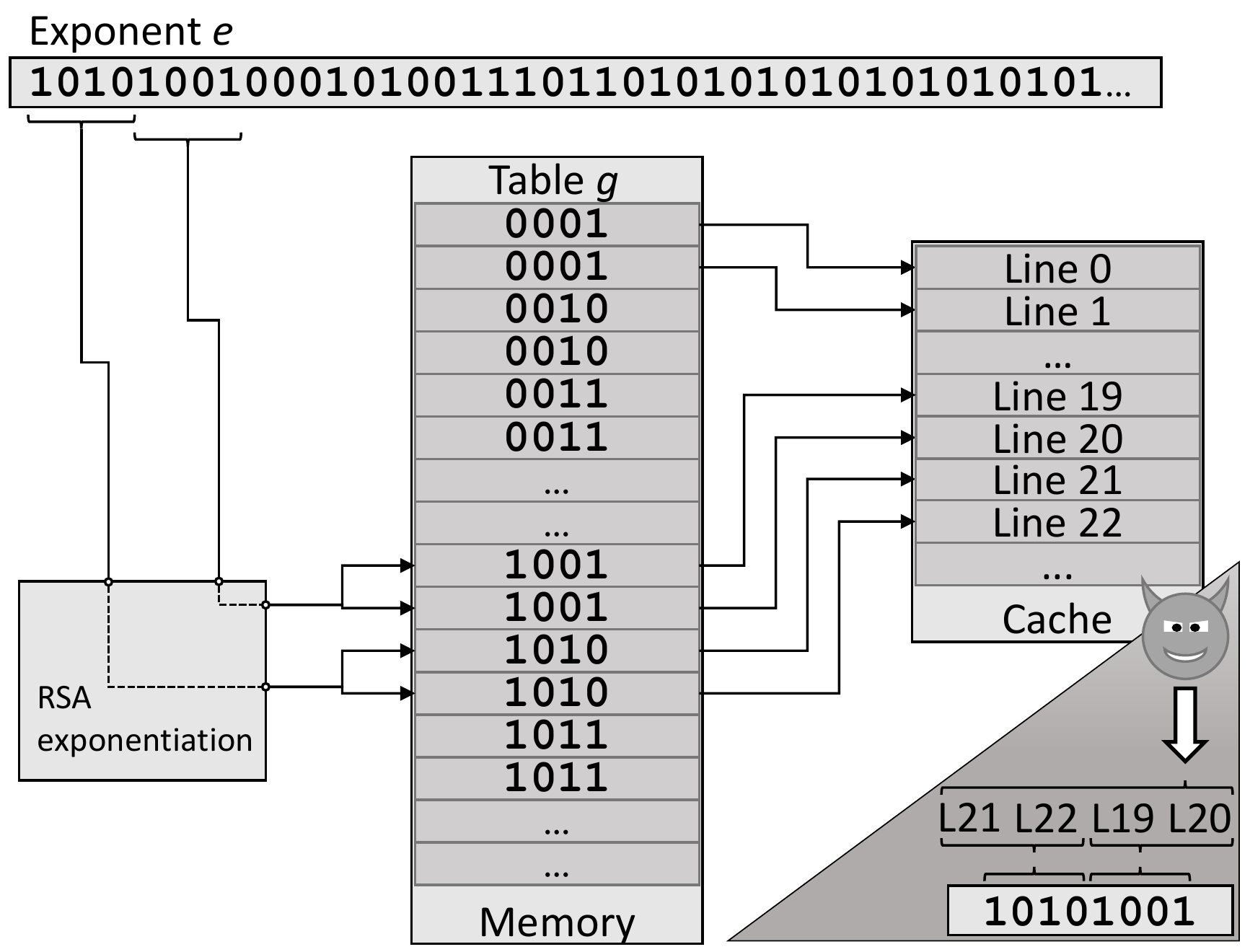}
	\caption{Memory accesses and cache updates in RSA exponentiation. The processed window value from exponent $e$ determines the accessed entry from table $g$ in memory which defines the updated cache line.}
	\label{fig:rsa}
	\vspace*{-.4cm}
\end{figure}

A pseudo code of the targeted exponentiation algorithm is shown in Algorithm~\ref{alg:exp}. Inputs of the algorithm are the base value $a$, the exponent $e$ (when CRT is used $d_p$ or $d_q$), and the public parameter $N$. The first step of the algorithm is a pre-computation of a multiplier table $g$ from the base value $a$. After that a $2^k$ representation of the exponent $e$ is computed, i.e., the exponent is divided into $\lceil n/k \rceil$ windows $(e_j, e_{j-1}, \ldots, e_1, e_0)$ of fixed size $k$ bits each. The algorithm iterates over all exponent windows starting from the most significant window (line 4 in Algorithm~\ref{alg:exp}) and, depending on the window value, it may perform a multiplication with a value from the multiplier table $g$. The value of the exponent window determines which pre-computed multiplier is accessed from the table $g$ on each iteration (line 7). Figure~\ref{fig:rsa} illustrates memory accesses and cache updates in the algorithm.

%

We compiled the RSA decryption implementation as an enclave with default optimization flags and compiler settings. When started, the enclave decrypts a single encrypted message. The private key was randomly chosen.

\begin{figure*}[th]
	\centering
	\includegraphics[trim={6cm 2.2cm 5cm 0.9cm}, clip, width=\textwidth]{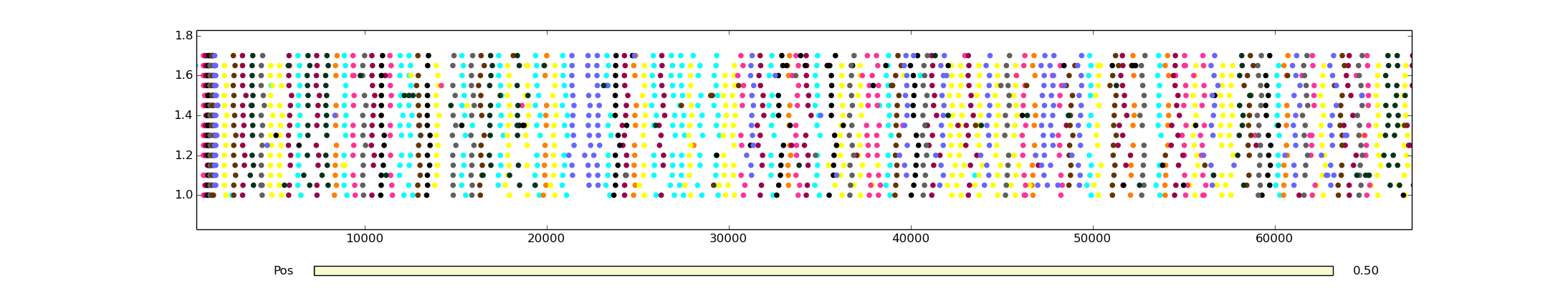}
	\caption{Access patterns of RSA key multipliers. Each dot represents 16 repeated memory accesses that correspond to a single multiplier in the precomputed table (see Algorithm~\ref{alg:exp}) and are observed from two monitored cache sets. We plot each monitored multiplier with a separate color. The monitoring process for each multiplier is repeated 15 times and each horizontal row in the plot represents one complete monitoring round. Most multiplier accesses are clearly distinguishable as separate colored vertical lines.}
	\label{fig:multipliers}
	\vspace*{-.2cm}
\end{figure*}

\subsection{Attack Details}
\label{subsec:attack-details}

Our attack proceeds as follows. Using the attack techniques described in Section~\ref{sec:design}, we monitor a single multiplier access at a time. Because each pre-computed multiplier is 1024 bits, this memory range corresponds to two cache sets.  We probe two monitored cache sets every $c$ cycles and divide the observed memory accesses into epochs of $p$ probes. Because each multiplier in the table is 1024 bits, accessing the multiplier causes 16 repeated memory accesses to the memory range of the table entry. If we observe 16 repeated accesses within one epoch, we mark the multiplier as a potential access candidate. We repeat this process for a subset of all possible multipliers (10 out of 16 in our case), because extracting a sufficiently large fraction of the key bits is enough to derive the entire key. We also observed significant cache access interference in some of the monitored cache sets\footnote{Presumably caused by the victim}, and therefore we opted not to monitor them. Finally, we repeat the entire process $t$ times.

Through experiments we observed that monitoring every $c=500$ cycles and dividing the monitoring into epochs of $p=33$ probes gave accurate results. To extract sufficiently large fraction of the key we needed to repeat the process $t=15$ times. Monitoring more than one multiplier at a time decreased multiplier access detection accuracy significantly. Similarly, performing monitoring more often than every $c=500$ cycles caused significant noise in measurements. The monitoring epoch $p=33$ probes was determined by the average execution time of a single exponentiation iteration.

\subsection{Attack Results}
\label{subsec:rsa-results}

Figure~\ref{fig:multipliers} shows our results on extracting the accessed pre-computed multipliers which in turn determine the private key. Each colored dot represents a multiplier access candidate. We plot different candidates with a separate color. Each horizontal row in the plot represents one complete monitoring round, where the monitoring process is performed separately for each multiplier (two cache sets). Because the entire monitoring process is repeated $t=15$ times, the plot has 15 horizontal lines. As can be seen from Figure~\ref{fig:multipliers}, most multiplier accesses are clearly distinguishable as colored vertical lines. 

To recover the multiplier access pattern, we analyze this plot manually. We use a simple heuristic of determining an access: if more than half of the monitoring rounds have the same value for the same epoch, we consider this value the accessed multiplier. If we observe no multiplier accesses in one epoch, the then we conclude that the exponent window for this iteration of exponentiation was zero (line 6 in Algorithm~\ref{alg:exp}). 

From the multipliers we construct a key candidate and compare it to the private key used by the enclave. Our attack extracts 70\% of the key bits correctly. This matches with the fraction of monitored cache sets $\frac{10+1}{16}=0.69$, where the $+1$ comes from the fact that the exponent windows value zero we learn without monitoring. From the extracted key bits, the complete private key can be efficiently recovered \cite{boneh-98}.

The closest previous cache attack is by Liu et al.~\cite{Liu2015}.\footnote{Percival~\cite{Per2005} demonstrates an attack against CRT RSA using sliding window on L1 cache, but does not report the number of decryptions.} They attack a sliding window RSA on through the Last Level Cache (LLC), because the attacker and the victim are running in different VMs. They are able to extract they key with tens of thousands of repeated decryptions, while we need 300 decryptions (10 observed multipliers, 15 repetitions, and two exponents). Although these two attack scenarios are not directly comparable, they do demonstrate that cache-based side-channel vulnerabilities are more severe in the SGX attacker model.



\newcommand{\kmer}{$k\textrm{-mer}$\xspace}
\newcommand{\kmers}{$k\textrm{-mers}$\xspace}
\newcommand{\twomer}{$2\textrm{-mer}$\xspace}
\newcommand{\twomers}{$2\textrm{-mers}$\xspace}
\newcommand{\fourmer}{$4\textrm{-mer}$\xspace}
\newcommand{\fourmers}{$4\textrm{-mers}$\xspace}

\section{Genomic Data Processing Attack}
\label{sec:genome}

In this section we describe our second side-channel attack on a genome data processing enclave.
Genome data processing is an emerging field that highly benefits from cloud computing due to the large amounts of data being processed.
At the same time, genome data is highly sensitive, as it may allow the identification of persons and carry information whether a person is predisposed to a specific disease.
Thus, maintaining the confidentiality of genomic data is paramount, in particular when processed in untrusted cloud environments.

In the remainder of the section we first introduce the general concept of the genome processing algorithm we used. Then, we describe the implementation of the algorithm on SGX, followed by attack details and our results.

Genome processing algorithms are just a representative for a large class of algorithms that produce memory accesses based on sensitive data, as we discuss in more detail in \cref{sec:discussion}.


%
%

\subsection{Victim Enclave}
\label{sec:genome:highlevel}

Genome sequences analysis is an important technique to identify individuals, persons or animals. 
By locating particular sequences in different location of a genome individuals can be distinguished.
Genome sequences are represented by the order of the four nucleotides adenine, cytosine, guanine and thymine, usually abbreviated by their first letter (A, C, G, T). 
Microsatellites, i.e., repetitive nucleotides base sequences, are commonly used for identifying individuals.
They usually range from two to five base pairs, occurring five to 50 times in a row in the genome.

Efficient search of large genome sequences is vital for these analysis methods.
Therefore, the data are usually preprocessed before the actual analysis is performed.
One common way of preprocessing is to divide the genome sequence into substrings of a fixed length $k$, called \emph{\kmer}. 
The \kmers represent a sliding window over the input string of genome bases.

\begin{figure}[tb]
	\centering
	\includegraphics[width=0.9\linewidth]{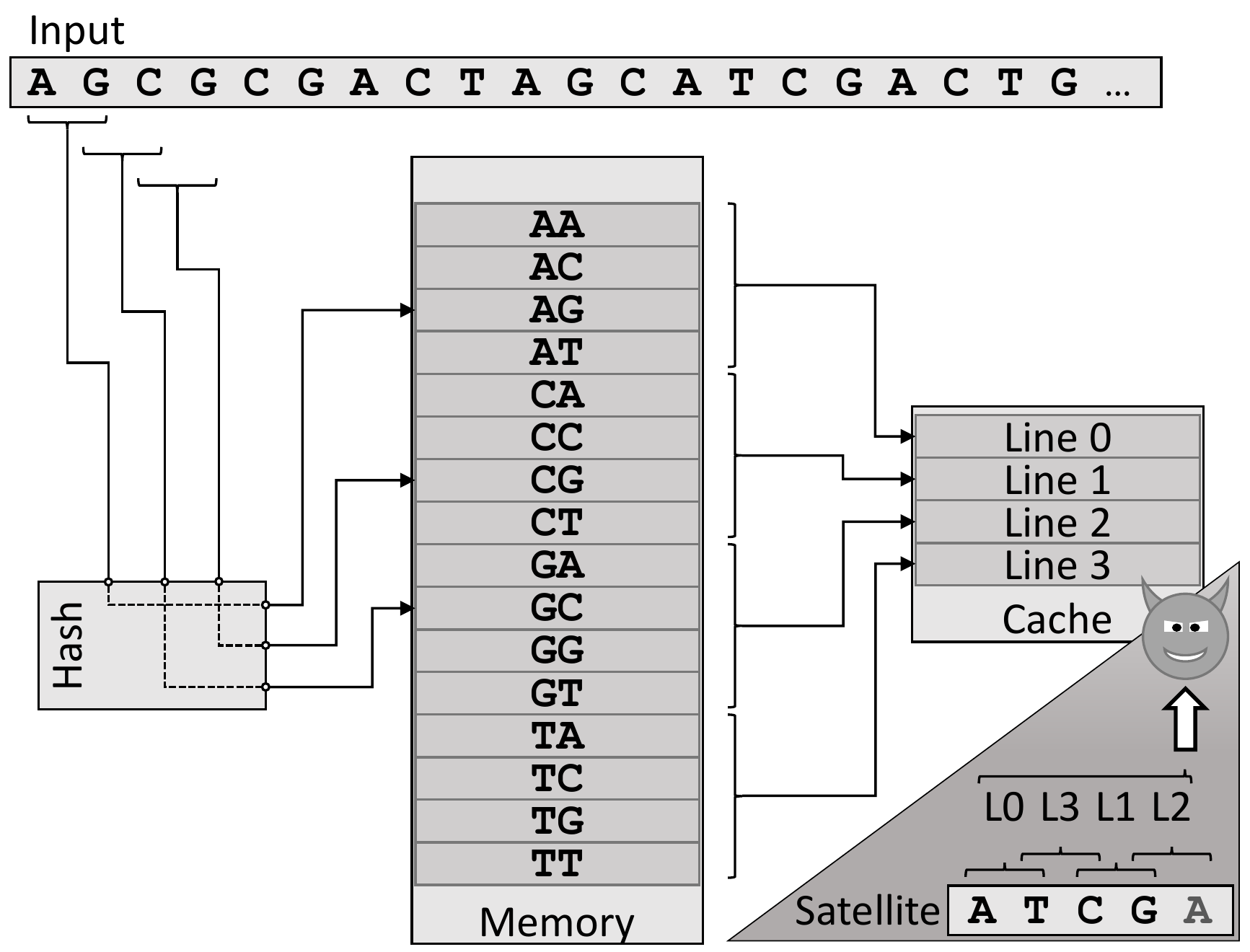}
	\caption{Genome sequence analysis based on hash tables; subsequences of the genome (called \kmers) are inserted into a hash table for statistical analysis and fast search for \kmers.}
	\label{fig:hashtable}
	\vspace*{-.4cm}
\end{figure}

In \cref{fig:hashtable} the input \texttt{AGCGC$\dots$} is split into \twomers. 
Starting from the left the first is \texttt{AG}, next the sliding window is moved by one character resulting in the second \twomer \texttt{GC}, and so on.

The \kmers are inserted into a hash table, usually, for each \kmer its position in the genome sequence is stored in the hash table.
Thus, given a \kmer that is part of a microsatellite one can quickly lookup at which position it appears in the input genome sequence.

Another use case is statistics of the input genome sequence, for instance, the distribution of \kmers in the sequence can easily be extracted from the hash table.

\paragraph{Primex.} Our victim enclave implements the preprocessing step for a genome sequence analysis algorithm, as described above.
We used and open source implementation of \kmer analysis tool called PRIMEX~\cite{primex}.\footnote{\url{https://www.researchgate.net/publication/233734306_mex-099tar}}
The tool inserts each \kmer into the hash table. Each hash table entry holds a pointer to an array, which is used to store the positions of each \kmer.

\subsection{Attack Details}

\algnewcommand\algorithmicswitch{\textbf{switch}}
\algnewcommand\algorithmiccase{\textbf{case}}
\algnewcommand\algorithmicassert{\texttt{assert}}
\algnewcommand\Assert[1]{\State \algorithmicassert(#1)}%
\algdef{SE}[SWITCH]{Switch}{EndSwitch}[1]{\algorithmicswitch\ #1\ \algorithmicdo}{\algorithmicend\ \algorithmicswitch}%
\algdef{SE}[CASE]{Case}{EndCase}[1]{\algorithmiccase\ #1}{\algorithmicend\ \algorithmiccase}%
\algtext*{EndSwitch}%
\algtext*{EndCase}%

\begin{algorithm}[tb]
	\caption{Hash-Index Generation}
	\label{alg:idxgen}
	
	\algorithmicrequire{Genome G with $\text{G}_i \in \{A, C, G, T\}$, $k \in \mathbb{N}_{> 0}$}
		\newline
		\algorithmicensure{Hash-Index H}
		
		\begin{algorithmic}[1]
			
			\State Let H $\gets$ HashTable with $4^k$ entries
			\For {\textbf{each } $k$-mer M $\in \text{G}$}
			\State Let $pos$ be the offset of M in G
			\State Let $idx \gets 0$
			\For {\textbf{each} nucleotide $n$ $\in$ M}
			
			\Switch{$p$}
			\Case{A: } $\overset{\sim}{n} \gets 0$ \EndCase
			\Case{C: } $\overset{\sim}{n} \gets 1$ \EndCase
			\Case{G: } $\overset{\sim}{n} \gets 2$ \EndCase
			\Case{T: } $\overset{\sim}{n} \gets 3$ \EndCase
			\EndSwitch
			
			\State $idx \gets 4 \cdot idx + \overset{\sim}{n}$
			\EndFor
			\State H[$idx$].append($pos$)
			\EndFor			
		\end{algorithmic}
\end{algorithm}

Our attack aims at detecting whether a specific subsequence, or microsatellite, is contained in the input genome sequence processed by the victim enclave. 
The microsatellite's position in the genome is revealed by the point in time when it is observed.
Due to the controlled environment of our attack the execution time of the victim if very deterministic, allowing precise positioning of the observation within the input sequence.
Additionally, the attack can be repeated for different microsatellites which allows the identification of individuals.

Through our cache side channel we can observe cache activities that can be linked to the victim's insertion operation into the hash table (\cref{alg:idxgen}).
\cref{fig:hashtable} shows that insertions into the hash table effect different cache lines.
For each \kmer the victim looks up a pointer to the associated array from the hash table. 
From the source code we learn the hash function used to determine the table index for each \kmer, by reversing this mapping we can infer the input based on the accessed table index.

Unfortunately, individual table entries do not map to unique cache lines. 
Multiple table entries fit within one cache line, so from observing the cache line accesses we cannot directly conclude which index was accessed.
This problem is illustrated in \cref{fig:hashtable}.
Here four table indexes map to a single cache line.
When the attacker observes the eviction of cache line $0$ (meaning it was accessed by the victim), it does not learn the exact table index of the inserted \kmer, but a set of candidate \kmers that could have been inserted (\texttt{\{AA,AC,AG,AT\}}).

However, the attacker can split up the microsatellite he is interested in into \kmers and determine which cache lines will be used when it appears in the input sequence.
In \cref{fig:hashtable} the microsatellite is split into four \twomers, where the first \twomer (\texttt{AT}) will be inserted in the first quarter of the table, hence, cache line 0 (L0) will be used by the victim enclave. 
The second \twomer (\texttt{TC}) will be inserted into the last quarter of the hash table, thus activating cache line 3 (L3).
Following this scheme the attacker determines a sequence of cache lines which will reveal to her that the microsatellite sequence was processed by the enclave.

\subsection{Attack Results}

We provided a real genome sequence string to the victim enclave and run it in parallel to our \primeandprobe attack code.
We chose $k = 4$ for the \kmers leading to $4^4 = 256$ \fourmers (four nucleotides possible for each of the four position).
Each \fourmer is represented by a unique table entry, each table entry is a pointer ($8\,byte$), and thus each cache line contains $64\,byte / 8\,byte = 8$ table entries.

In our attack we were searching for a tetra-nucleotide microsatellite of length ten ($(\texttt{ATCG})_{10}$).
First, the four \fourmers occurring repeatedly in microsatellite are determined, and for each \fourmer the corresponding cache lines: $\texttt{ATCG} \Rightarrow \mathrm{cache\ line\ }62$; $\texttt{TCGA} \Rightarrow \mathrm{cache\ line\ }63$; $\texttt{CGAT} \Rightarrow \mathrm{cache\ line\ }22$; $\texttt{GATC} \Rightarrow \mathrm{cache\ line\ }39$.

\begin{figure*}[th]
	\centering
	\includegraphics[trim={6.6cm 2.4cm 5.8cm 1.15cm}, clip, width=\textwidth]{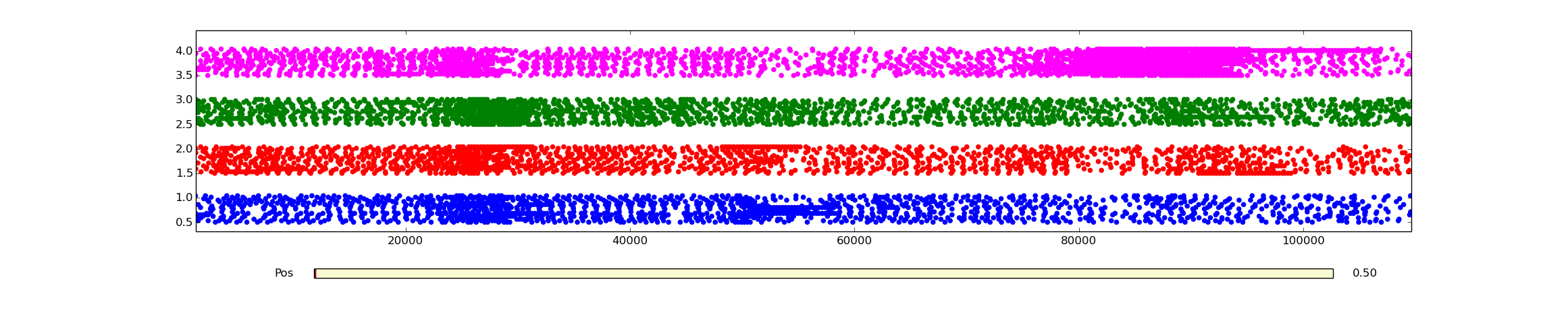}
	\caption{Access pattern of hash table accesses by PRIMEX processing a genome sequence~\cite{primex}. Four cache sets are shown in different colors with 20 repeated measured for each cache set. The cache sets are correspond to the \fourmers of the microsatellite \texttt{ATCG}. At $x \approx 25,000$ increased activity in all four cache sets indicates the occurrence of the microsatellite in the processed genome sequence.}
	\label{fig:satellite}
	\vspace*{-.2cm}
\end{figure*}

We monitor these four cache lines individually and align them, as shown in \cref{fig:satellite}.
When the microsatellite appears in the input string, the cache lines $62$, $63$, $22$ and $39$ will all be used repeatedly by the victim enclave.
This increase in utilization of these cache sets can be observed in the measurements.
In \cref{fig:satellite} at $x \approx 25,000$ the increased density of observed cache events is visible.
Since all four cache lines are active at the same time, one can conclude that the microsatellite did occur in the input sequence.

\paragraph{False positive analysis.} False positives can occur for two reasons, (1) sequences that map to the same cache lines as the microsatellite we are searching for, (2) noise in the cache.
We calculated the set of accessed cache lines for all possible tetra-nucleotide microsatellite, we found no collusions.
The only exception is are \fourmers that are one of the three possible rotations of the microsatellite sequence we are searching for.
This means that no other sequence of \fourmers produces activity in the same sets of cache lines and cause a false positive.

False positives due to noise are very unlikely due to the fact that we are observing four cache lines.
\cref{fig:satellite} shows extensive activation in the top cache line (pink) in the interval $x \approx 80,000$ to $x \approx 95,000$.
However, in all three other cache lines there is low activity making this event clearly distinguishable from a true positive event.

\section{Countermeasure Analysis}
\label{sec:countermeasures}
In this section we discuss potential countermeasures against cache-based side channel attacks and elaborate on their applicability to protection of SGX enclaves. 


\paragraph{Cache disabling.}
The most straightforward countermeasure against cache-based side channels is to disable caching entirely~\cite{Aciicmez2010}. This approach, however, defeats performance optimizations for which cache memory was intended for at first place, resulting in severe performance degradation. More fine-grained approach is to disable the cache only when security critical code is scheduled for execution. In the context of SGX, it would mean to disable caching during enclave execution, which may still be prohibitively expensive given that SGX enclaves may need to process large datasets (e.g., human DNA) or perform expensive computation (e.g., cryptography), or run large applications. For instance, Haven architecture~\cite{baumann_shielding_2014} loads the entire database management system (DBMS) into an enclave. 

\paragraph{Architectural changes to cache organization.} Other approaches proposed to mitigate cache-based side channels with low overhead through redesign of the cache hardware. The first line of works includes proposals of new cache designs by applying the idea of cache partitioning so that security sensitive code never shares cache memory with untrusted processes (e.g., \cite{Page2003,Page05partitionedcache,WangLee2006,WangLee2007,DoJaLoAbPo2012}), while another one concentrates on access randomization within cache memory~\cite{WangLee2007,WangLee2008,Keramidas2008,LiLe2014}. However, these approaches would require a radical change to current cache designs, which cannot be easily implemented in practice.  In particular, Intel processors with SGX extensions do not implement any countermeasures against cache side-channel attacks at the architectural level. 

Sanctum~\cite{sanctum} flushes the L1 cache on switches between enclave and non-enclave mode.
This approach does not stop our attack since our attack runs in parallel to the enclave.
The enclave is not interrupted to probe the cache, and hence, no mode switch and no cache flushing is triggered.

\paragraph{Obfuscation techniques.} \label{sec:countermeasures:oblivious}
The state-of-the-art obfuscation technique to defeat information leakage via side channels is Oblivious RAM (ORAM) \cite{GolOst1996,stefanov-ccs13,permuteRAM}, which provides means to hide memory access patterns of programs by continuously shuffling and re-encrypting data as they are accessed in RAM memory, disk or from a remote server. ORAM is typically applied in server-client models, and requires the client to store some state that
is updated throughout the subsequent execution. While one could think of using similar techniques for cache protection, they are not directly applicable, as it is challenging to store ORAM internal state securely. Without hardware support this would require storing client state in a cache side-channel oblivious way, which is unfeasible given small size of every cache line.

Other obfuscation techniques suggest to perform periodic scrubbing and flushes of shared caches~\cite{Zhang2013} or add noise to memory accesses~\cite{Page2003,Osv2006} to interfere with the signal observable by the attacker. These techniques, however, introduce a significant overhead and will not necessarily eliminate the attack we presented. Especially, these countermeasures are less effective on systems supporting simultaneous multithreading, where two threads or processes can be executed literally simultaneous, not in a time-sharing fashion. In this case the attacker process running in parallel with the victim can still observe memory access patterns between scrubbing and flushing rounds. Furthermore, an attacker may collect multiple execution traces and process them to filter out the injected noise.

%
 %
%



%
%


\paragraph{Application-level hardening.} \label{sec:countermeasures:hardening}  
Application-level hardening techniques modify applications in order to protect their secrets from side-channel leakage. 
Such solutions can be classified into two categories: (i)~Side-channel free implementations (e.g., for cryptographic algorithms AES and RSA~\cite{BrGrSe2006,Konighofer2008}) and (ii) automated tools that can be applied to existing programs and do not require manual program modification~\cite{CoVeBoSu2009,Cleemput2012,CHBLF2015}. 
However, side-channel free implementations are application-specific and require significant manual effort and thorough understanding of the subject matter, although generally application developers cannot be expected to be security experts. On another hand, approaches that rely on automated processing, e.g., compiler transformations for limiting branching on sensitive data~\cite{CoVeBoSu2009} or reducing/masking timing variability~\cite{Cleemput2012,CHBLF2015}, typically cannot eliminate side channels entirely, since opportunities to do so automatically are limited.


In context of SGX, Shinde et al.~\cite{ShChNaSa2016} proposed application hardening as a mitigation technique against page fault based side channel. The solution relies on developer-assisted compiler optimizations, or, if applied generally, imposes a performance overhead up to 4000x. While similar approach can be used to defeat cache-based side channels, associated drawbacks (either manual effort or impact on performance) limit its practicality.




\paragraph{Randomization.} \label{sec:countermeasures:randomization}
Address Space Layout Randomization (ASLR)~\cite{PaX-ASLR} is another alternative, which might provide a viable solution against cache-based side channels. Despite the fact that it was designed as a defense mechanism against code reuse attacks, similarly to ORAM, it can hide access patterns to secret-dependent code and data, if applied to randomize enclave's memory layout. 

ASLR randomizes the \emph{base addresses} of loaded code and data in memory, making memory layout of the vulnerable process different across different instances and even across different runs. In this form, ASLR is deployed on most mainstream computing platforms for PCs and mobile devices, including Windows, Linux, iOS and Android. However, in recent years there were many attacks demonstrated that have shown that randomization of base addresses provides insufficient entropy and can be brute forced~\cite{ShGoMoPfBo2004,LiJiDeJiDa2011}, or information about them can be obtained via information leakage attacks, e.g., by exploiting information used for linking dynamic libraries~\cite{acsac09:saratoga} or exploiting information leakage bugs (e.g.,~\cite{teso2001}). These attacks motivated further development of more fine-grained memory randomization forms, which randomize application binaries at granularity of functions~\cite{KiJuBoXuNi2006}, basic blocks\footnote{A basic block is a sequence of machine instructions with a single entry and exit instruction, where the latter one
can be any branch instruction the processor supports}~\cite{WaMoHaLi2012,TUD-CS-2013-0042} and even single instructions~\cite{PaPoKe2012,Hi2012}. 

Fine-grained memory randomization techniques was undermined by Snow et al.~\cite{TUD-CS-2013-0026}, who demonstrated a dynamic code reuse attack that could disclose memory layout of the victim application through repeated exploitation of the memory leakage vulnerability and constructing the attack payload at the time when the attack is executed. Doing so requires certain amount of time, which motivated new approaches that consider periodic re-randomization performed at runtime~\cite{LNBL2016}.




Recently, Seo et al.~\cite{SGXShield2017} proposed SGX Shield framework that enables code randomization for SGX enclaves. While the primary goal of SGX Shield is to protect enclaves from exploitable software bugs, authors mention that randomization imposes additional burden to side channel attackers, and in particular it provides reasonable protection against page-fault side-channel attacks, as it forces an attacker to brute force $2^{7}$ times in order to identify a single address value. However, this argumentation does not directly apply to the case of cache-based side channels, because SGX Shield concentrates on randomization of code, but does not randomize data. Hence, SGX shield cannot hide data-dependent memory access patterns. On another hand, randomization of data segments is challenging due to dynamic data allocations, large data objects (e.g., tables) that need to be split up and randomized, and pointer arithmetic which is typically used to access parts of large data objects (e.g., base-pointer offsets are often used to access table entries). 






\paragraph{Attack detection.} \label{sec:countermeasures:detection}
Recently, two interesting works proposed detection methods for side-channel attacks that are based on frequent interruption of the victim enclave~\cite{t-sgx,incognito2017}.
In particular, both solutions aim at mitigating side channels based on page-faults~\cite{Xu2015}.
Here the OS incurs page faults during enclave execution and learns the execution flow of the enclave from the requested pages. In particular, both works suggest using a hardware implementation of transactional memory in Intel processors called Intel Transactional Synchronization Extensions (TSX) to notify an enclave about a (page fault) exception without interference by the system software.
This generally enables enclaves to detect if their execution was preempted or interrupted. 
D{\'e}j{\'a} Vu~\cite{incognito2017} also aims at defeating cache-based side-channel attacks that preempt the victim enclave frequently to more accurately observe the victim's cache accesses.
However, as we show in our work, cache-based side channels do not necessarily require preemption of the protected application to make side channel observations.
Hence, such countermeasures cannot defeat our attack.

\paragraph{Summary.} 
We believe that system-level defense mechanisms like memory randomization are more plausible, as they provide protection to any program, independently if they were implemented by security experts, and are more effective in closing side channels entirely. They do not require changes to underlying hardware and impose moderate performance overhead. However, the only memory randomization solution for SGX enclaves SGX Shield~\cite{SGXShield2017} does not support randomization of data objects, which is challenging to achieve, as we elaborated above. We aim to explore possible designs and ways to overcome associated challenges in our future work.

\section{Discussion}
\label{sec:discussion}

\paragraph{Other algorithms.} In his paper we have demonstrated information leakage through secret-dependent data accesses in RSA decryption and human genome indexing. Both of these target algorithms construct a table that is repeatedly accessed while the algorithm processes through the confidential data. The same high-level algorithmic pattern is not limited to these two applications, but also found in many other domains, such as database indexing, compression algorithms, image processing. Based on our results, there is reason to believe that many of these algorithms would be vulnerable to cache-based information leakage, but we leave the demonstration of practical attacks as future work.

\paragraph{Lessons learned.} 
Through our experiments we observed that there are certain key factors that determine how vulnerable a particular algorithm is to cache-based information leakage. The size of the constructed table determines if, and how many, multiple table entries map to the same cache set, and thus cause increased cache monitoring interference. The frequency of table accesses defines the available time budget for monitoring on each algorithm iteration round, and thus the probability of catching the data access. Large table entries and repeating patterns in the processed confidential data cause repeated data accesses that make the algorithm (and data) more vulnerable to our attacks.



\section{Related Work} \label{sec:relwork}

In this section we review works related to the Intel SGX supported applications, to side channel attacks mounted against SGX enclaves and to cache-based side channel attacks on non-SGX platforms. 

\paragraph{SGX applications.} 
First applications leveraging SGX support were already developed and consider cloud scenarios~\cite{baumann_shielding_2014,ScCoGkPeMaRu2015,DiSaChOoZh2015,DeaGhe2008,HuZhXuPeWi2016,YSDCMOONF2009} and beyond~\cite{KiShHaKiHa2015,Shih2016}. All these applications are potential targets to cache-based side-channel attacks, and if not designed to be side-channel resistant, they may leak application secrets in the similar way as the genome processing application which we investigated in this paper (cf.~Section~\ref{sec:genome}). 

\paragraph{Side-channel attacks on SGX.}
The SGX architecture was analyzed by Costan and  Devadas~\cite{Cos2016}, who mentioned that SGX is likely to be vulnerable to side-channel attacks, that could potentially be used to leak protected secrets from within the SGX enclaves. Xu et al.~\cite{Xu2015} demonstrated page-fault based side-channel attacks on SGX, where an untrusted operating system infiltrates secrets from protected applications by tracking memory accesses at the granularity of memory pages. While cache-based side channel attacks, which we study in this paper, generally achieve more precise tracking of memory accesses at the granularity of cache lines, they have not been investigated in context of SGX in previous works.

\paragraph{Cache attacks.}
The first cache-based side channel attack~\cite{Per2005} demonstrated information leakage via L1 cache and was successfully applied to reveal RSA keys of OpenSSL implementation through monitoring accesses to the table with precomputed multipliers, which are used by the algorithm throughout the exponentiation. 
Detailed performance comparison to this attack is not possible, as the paper does not report details, such as how many repetitions are needed to extract the key. The attack was performed on more than 10 years old platform.

The side-channel free implementation of RSA was proposed by Brickell et al.~\cite{BrGrSe2006}. It relies on a technique called \emph{scatter-gather} to interleave the multipliers in memory, which ensures that the same cache lines are accessed irrespective of the multiplier. However, eventually memory accesses within the same cache line with different offsets may also have time variations~\cite{OptManual2012}.  
This was exploited by CacheBleed attack~\cite{CacheBleed2016}, successfully recovering 60\% of exponent bits of the RSA key after observing 16,000 decryptions.
We hypothesize that side channel attack based on cache-bank conflicts may also be applied to SGX enclaves, although we have not investigated this aspect in our work.


Osvik et al.~\cite{Osv2006} formalized two cache-based side channel attack techniques, \emph{Evict+Time} and \emph{Prime+Probe}, which since then have been used to attack various cryptographic implementations~\cite{NeSeWa2006,OsShTr2006}, were applied to last level cache and used to build cross-core side channels~\cite{IrEiSu2015,Liu2015}. Furthermore, they were also shown to be applicable to mobile and embedded platforms~\cite{BoEiPaWi2010,WeHeSt2012,SpPl2013,SpGe2014}. In a context of cross-core attacks, new and more complex attack techniques were developed, such as Flush+Reload~\cite{YarFal2014}, Evict+Reload~\cite{GrSpMa2015}, and Flush+Flush~\cite{MarWag2015}. Similarly to us, some of the cross-core attacks~\cite{Liu2015} target RSA decryption. These attacks tens of thousands of repetitions, while our attack requires only about 300 executions.

Uhsadel et al.~\cite{UGV08} study the use of hardware performance counters (HPCs) for side-channel attacks.
They use HPCs to observe the behavior of their victim directly, e.g., record cache hit/miss events of the victim.
This approach is not suitable for SGX enclaves because enclaves do not update HPCs.
In contrast, we use HPCs to record cache events of the attacker's \primeandprobe code.

\section{Conclusion} 
\label{sec:conclusion}

Researchers have assumed that SGX may be vulnerable to cache-based information leakage. However, before our work, the practicality and the extent of such leakage was not well understood. In this paper we have demonstrated that cache attacks on SGX are indeed a serious concern. Our goal was to develop an attack that cannot be mitigated by the known countermeasures, and therefore we mount the attack on uninterrupted enclave execution. Such attack approach involves technical challenges. To address them, we developed a set of novel noise reduction techniques. We demonstrated them on RSA decryption and human genome indexing. Our attacks are more efficient than previous cache attacks and harder to mitigate than previous SGX side-channel attacks.

\bibliography{bibliography}{}
\bibliographystyle{abbrv}

%

\end{document}